\documentclass[11pt]{article}
\usepackage{epsfig,amsmath,amsthm,graphicx,amssymb,overpic}
\usepackage{epsfig,amsmath,graphicx,amssymb,overpic,cite}
\usepackage{listings,diagbox,color}

\def\be{\begin{equation}}
\def\ee{\end{equation}}
\def\bee{\begin{eqnarray}}
\def\ene{\end{eqnarray}}
\def\bes{\begin{subequations}}
\def\ees{\end{subequations}}

\def\v{\vspace{0.1in}}

\newcommand{\bp}{{\bf p}}

\newcommand{\bx}{{\bf x}}

\begin{document}

\baselineskip=12pt
\renewcommand {\thefootnote}{\dag}
\renewcommand {\thefootnote}{\ddag}
\renewcommand {\thefootnote}{ }

\pagestyle{plain}

\begin{center}
\baselineskip=15pt \leftline{} \vspace{-.3in} {\Large \bf
 Deep neural networks for solving forward and inverse problems
 of  (2+1)-dimensional nonlinear wave equations with rational solitons} \\[0.2in]
\end{center}

\begin{center}
Zijian Zhou$^{1,2}$, Li Wang$^{3,4}$, and Zhenya Yan$^{1,2,*}$\footnote{$^{*}${\it Email address}: zyyan@mmrc.iss.ac.cn (Corresponding author)}  \\[0.05in]
{\it \small $^{1}$Key Laboratory of Mathematics Mechanization, Academy of Mathematics and Systems Science, \\ Chinese Academy of Sciences, Beijing 100190, China \\
 $^{2}$School of Mathematical Sciences, University of Chinese Academy of Sciences, Beijing 100049, China\\
 $^{3}$Yanqi Lake Beijing Institute of Mathematical Sciences and Applications, Beijing, 101408, China \\
  $^{4}$Yau Mathematical Sciences Center and Department of Mathematics, Tsinghua University, Beijing, 100084, China} \\
\end{center}


{\baselineskip=13pt



\noindent {\bf Abstract.}\, In this paper, we investigate the forward problems on the data-driven rational solitons for the (2+1)-dimensional KP-I equation and spin-nonlinear Schr\"odinger (spin-NLS) equation via the deep neural networks leaning. Moreover, the inverse problems of the (2+1)-dimensional KP-I equation and spin-NLS equation are studied via deep learning. The main idea of the data-driven forward and inverse problems is to use the deep neural networks with the activation function to approximate the solutions of the considered (2+1)-dimensional nonlinear wave equations by optimizing the chosen loss functions related to the considered nonlinear wave equations.


\vspace{0.1in} \noindent {\it Keywords:} (2+1)-dimension nonlinear wave equations; deep neural networks learning; activation function; data-driven rational solitons;  data-driven parameter discovery



\section{Introduction}

In the past decade, data analysis and machine learning with neural networks have been paid more attention to and 
achieved some significant advances due to the explosive development of big data and computing resources. As a result,
these progresses further promoted the rapid development of other related fields, including computer vision, natural language processing, cognitive science, optical text/character recognition and data assimilation~\cite{DL1,DL2,DL3,DL4,DL5,DL6}, etc..
Hornik {\it et al} showed that the standard multi-layer feedforward networks with one hidden layer and arbitrary bounded and nonconstant activation function could approximate any Borel measurable function in any accuracy if the sufficiently many hidden
units are available~\cite{NN1,NN2}. And the other some works provided some new perspectives on the functions of the neural networks (see, e.g., Refs.~\cite{NN5,NN6,NN7,NN8} and references therein). The main idea of deep neural networks  learning is to use the simple compositions of linear functions and  nonlinear activation functions to represent the solved functions.
Some recent studies have focused on the applications of deep learning in the high-dimensional problems, including the partial differential equations (PDEs) \cite{hdPDE1,hdPDE2}, stochastic differential equations (SDEs) \cite{hdPDE3,hdPDE4} and molecular dynamics \cite{hd1,hd2}.

In recent years, many deep neural networks focused on the study of differential equations, such as the physics-informed neural networks (PINNs)~\cite{raiss19,PINN1,PINN2,PINN4}, deep Ritz method~\cite{ritz}, deep Galerkin method (DGM) \cite{DGM}, PDE-net~\cite{PDEn1,PDEn2}, and etc. The physical constraints are added to the loss functions to powerfully learn the models~\cite{raiss19}. In Refs.~\cite{ritz,PINN1}, the equation loss was replaced by the variational loss. In addition, many other works extended these deep neural networks learning methods solving the PDEs (see, e.g., Refs.~\cite{PDE2,pde-c,PDE3,PDE4,PDE5,PDE6,PDE7,PDE8,raiss18}).

In this work, we would like to investigate the following forward and inverse problems of high-dimensional nonlinear wave equations via the deep learning method. We consider these problems in $(\bx, t)\in D\times[t_0,t_1]\in \mathbb{R}^{n+1}$.

\begin{itemize}

\item {Forward problem:} For the given higher-dimensional nonlinear equation ($F(q,q_{\bx},q_t,...)=0$) with the initial data ($q(\bx,0)=q_0(\bx)$) and periodic boundary conditions
    \bee\label{prob1}
    \left\{\begin{array}{l}
    F(q,q_{\bx},q_t,...)=0,\quad \vspace{0.1in}\\
    q(\bx,t_0)=q_0(\bx), \quad \vspace{0.1in}\\
    q(\bx_{B_1},t)=q(\bx_{B_2},t), \quad
    \end{array}\right.
    \bx\in D,\quad \bx_{B_1},\bx_{B_2}\in \partial D,\quad t\in [t_0, t_1]
    \ene
  one would like to use the deep learning method to emulate the data-driven solution $q(\bx,t)$ of Eq.~(\ref{prob1}).

\item {Inverse problem:} For the given solution data ($q_0(\bx,t)$), one would like to use the deep learning method to discover the unknown parameters $\bp\in {\mathbb R}^m$ of $F(\bp,q,q_x,q_t,...)=0$ by considering
    \bee\label{prob2}
    \left\{\begin{array}{l}
    F(\bp,q,q_{\bx},q_t,...)=0,\quad \vspace{0.1in}\\
    q(\bx,t)=q_0(\bx,t), \quad
    \end{array}\right.
    \bx\in D,\quad t\in [t_0, t_1], \quad \bp\in {\mathbb R}^m.
    \ene
where $\bx =(x_1,x_2,...,x_n)$ and $\bp=(p_1,p_2,...,p_m)$.

\end{itemize}

The rest of this paper is arranged as follows. In Sec. 2, we simply introduce the PINN scheme for the forward problem, and study
the data-driven rational solitons of the (2+1)-dimensional KP-I equation and spin-NLS equation via the PINN deep learning. In Sec. 3, we introduce the PINNs scheme for the inverse problem, and apply it to discover the parameters of the associated functions in the (2+1)-D KP-I equation and spin-NLS equation. Finally, we give some conclusions and discussions in Sec. 4.

\section{Forward problems in  high-dimensional nonlinear wave equations}

\subsection{The deep neural networks learning scheme}

In this subsection, we would like to briefly introduce the PINN deep learning method~\cite{raiss19} for the forward problem (see the data-driven solutions of Eq.~(\ref{prob1})) in the high-dimensional space. It is well-known that the curse of dimensionality
is a big problem in usual numerical algorithms, but the high-dimensional problem is not a big deal for the deep learning method. The main idea of the PINN method is to use a deep neural network to fit the solutions of Eq.~(\ref{prob1}). The number of input neuron denotes the number of variable. Fig.~(\ref{fig1-DNN}) depicts the structure of the PINN method for forward problem in $(2+1)$-
D case. Next, we will describe its structure in detail. For convenience, we illustrate the method in (2+1)-D case, and
just discuss the periodic boundary conditions. In fact, other kinds of boundary conditions can also be considered in the same way.

We assume that $q(\bx,t)$ is a solution of $F(q,q_x,q_t,...)=0$. Let $q(\bx,t)=u(\bx,t)+iv(\bx,t)$ with $u(\bx,t),\,v(\bx,t)$ being its real and imaginary parts, and $\bx$ represent the spatial variable (i.e. $\bx=(x_1,x_2) $). The complex-valued function $F(q,q_x,q_t,...)$ can be written as $F(q,q_x,q_t,...)=F_u(q,q_x,q_t,...)+iF_v(q,q_x,q_t,...)$ with $F_u(q,q_x,q_t,...),\, F_v(q,q_x,q_t,...)$ being its real and imaginary parts, respectively. We will approximate the solution $q(\bx,t)$ by using a complex-valued deep neural network $\hat q(\bx,t)=(\hat u(\bx,t),\, \hat v(\bx,t))$ written as
\begin{lstlisting}
def q(x1, x2,  t):
    q = neural_net(tf.concat([x1,x2,t],1), weights, biases)
    u = q[:,0:1]
    v = q[:,1:2]
    return u, v
\end{lstlisting}

\begin{figure}[!t]
\begin{center}
{\scalebox{0.75}[0.75]{\hspace{0.25in}\includegraphics{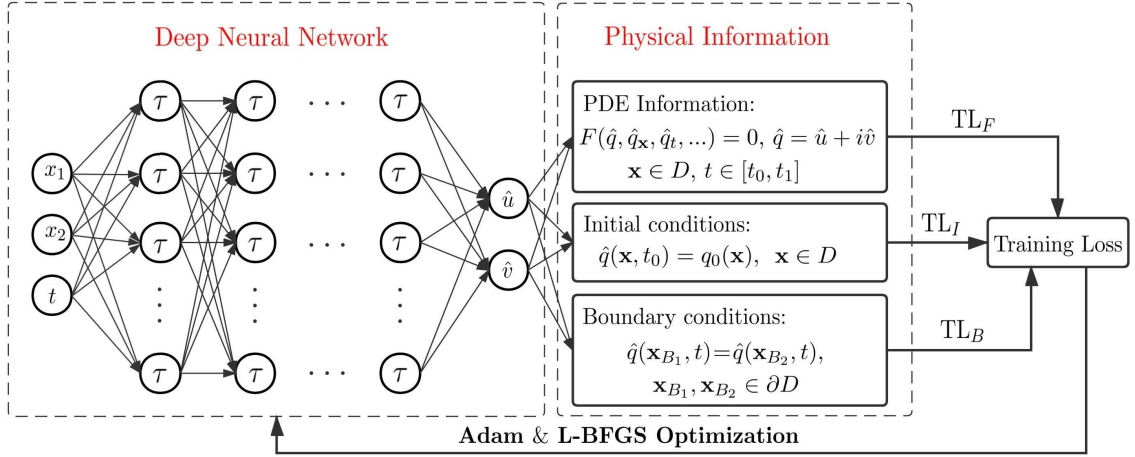}}}
\end{center}
\par
\vspace{0.05in}
\caption{\small The PINN scheme solving the forward problem (\ref{prob1}), where ${\cal T}$ denotes the activation function (e.g., ${\cal T}=\tanh(\cdot)$).}
\label{fig1-DNN}
\end{figure}

Based on the defined function $q(\bx,t)$, the physics-informed neural network $F(q,q_x,q_t,...)$ can be taken as
\begin{lstlisting}
 def F(x1, x2,  t):
     u, v = q(x1, x2,  t)
     u_t = tf.gradients(u, t)[0]
     u_x1 = tf.gradients(u, x1)[0]
     u_x2 = tf.gradients(u, x2)[0]
     ...
     v_t = tf.gradients(v, t)[0]
     v_x1 = tf.gradients(v, x1)[0]
     v_x2 = tf.gradients(v, x2)[0]
     ...
     F_u = (real part of F(q,q_x,q_t,...))
     F_v = (imaginary part of F(q,q_x,q_t,...))
     return F_u, F_v
\end{lstlisting}
where $\hat q(\bx,t)=\hat u(\bx,t) +i\hat v(\bx,t)$ and $F(q,q_x,q_t,...)=F_u(q,q_x,q_t,...)+ iF_v(q,q_x,q_t,...)$ share the same parameters, weights and biases, and the deep neural network can be learned by optimizing the training loss (TL). In the forward problem (\ref{prob1}), the training loss (TL) is assumed as the sum of three parts including the initial loss (${\rm TL}_{I}$), the boundary loss (${\rm TL}_{B}$), and the equation loss (${\rm TL}_{F}$):
\bee\label{loss}
  {\rm TL}={\rm TL}_{I}+{\rm TL}_{B}+{\rm TL}_{F},
\ene
where the three parts of loss are defined by the mean squared errors (i.e., $\mathbb{L}^2$-norm):
\bee\begin{array}{l}
 {\rm TL}_{I}=\displaystyle\frac{1}{N_I}\sum_{j=1}^{N_I}\left[\left(\hat u(\bx_I^j, t_0)-u(\bx_I^j,t_0)\right)^2+\left(\hat v(\bx_I^j, t_0)-v(\bx_I^j,t_0)\right)^2\right],\v\\
  {\rm TL}_{B}=\displaystyle\frac{1}{N_B}\sum_{j=1}^{N_B}\left[\left(\hat u(\bx_{B_1}^j, t_{B}^j)-\hat u(\bx_{B_2}^j, t_{B}^j)\right)^2+\left(\hat v(\bx_{B_1}^j, t_{B}^j)-\hat v(\bx_{B_2}^j, t_{B}^j)\right)^2\right],\v\\
  {\rm TL}_{F}=\displaystyle\frac{1}{N_F}\sum_{j=1}^{N_F}\left(F_u^2(\bx_F^j, t_F^j)+F_v^2(\bx_F^j, t_F^j)\right),
 \end{array}
\ene
with $\{\bx_I^j,\, u(\bx_j,t_0),\, v(\bx_j,t_0)\}_{j=1}^{N_I}$ denoting the initial data, $\{t_B^j,\, \hat u(\bx_{B_{1,2}}^j, t_{B}^j),\, \hat v(\bx_{B_{1,2}}^j, t_{B}^j)\}_{j=1}^{N_B}$ standing for the periodic boundary data ($\bx_{B_{1,2}}^j=(x_{1B_{1,2}}^j, x_{2B_{1,2}}^j)$ with $x_{iB_{1}}^j$, $x_{iB_{2}}^j$ representing the lower and upper bounds in $i$-axis), $\{\bx_F^j,\, t_F^j\}_{j=1}^{N_F}$ representing the collocation points of PINNs $F(q,q_x,q_t,...)=F_u(q,q_x,q_t,...)+iF_v(q,q_x,q_t,...)$ within a spatio-temporal region
$(\bx,t)\in [-L, L]^2\times (t_0, t_1]$. All of these sampling points are generated using a space filling Latin Hypercube Sampling strategy~\cite{Latin}.

We would like to choose a multi-layer fully-connected neural network with some neurons per layer and a hyperbolic tangent activation function $\tanh(\cdot)$. We assume that $Q^j=(a_1^j, a_2^j,..., a_{m_j}^j)^T$ and $B^j=(b_1^j, b_2^j,..., b_{m_j}^j)^T$ denote the output and  bias column vectors of the $j$-th layer, respectively, and $W^{j+1}=(w_{ks}^{j+1})_{m_{j+1}\times m_j}$ stands for the weight matrix of the $j$-th layer. In the input and output layers of neural networks, $Q^0=(\bx,t)^T=(x_1,x_2,t)^T$, $Q^{M+1}=(\hat u, \hat v)^T$. The calculation in each hidden layer is shown as follows:
\bee
\begin{array}{l}
  Q^{j+1}=\tanh\left(W^{j+1}A^{j}+B^{j+1}\right)\displaystyle\left(\tanh\left(\sum_{k=1}^{m_{j}} w_{1k}^{j+1} a_k^{j}+b_1^{j+1}\right),\cdots,
    \tanh\left(\sum_{k=1}^{m_{j}} w_{m_{j+1}k}^{j+1} a_k^{j}+b_{m_j}^{j+1}\right)\right)^T.
\end{array}
\ene
The real and imaginary parts of solution, $u(\bx,t)$ and $v(\bx,t)$, are approximated by the two outputs, $\hat u$ and $\hat v$, of one neural network, respectively (see Fig.~\ref{fig1-DNN}).

In the following, we consider the data-driven rational solitons of the (2+1)-dimensional KP-I equation and spin-NLS equation via the deep neural networks learning.



\subsection{Data-driven rational solitons of the (2+1)-D KP-I equation}

In this part, we would like to consider the data-driven rational solitons of (2+1)-dimensional KP-I equation~\cite{kp70}
 \bee\label{KP-I}
(q_t+6qq_x+q_{xxx})_x-3q_{yy}=0,
\ene
which can be used to describe the propagation of long ion-acoustic waves with the small amplitude in plasmas~\cite{kp70}, and the capillary gravitational waves on a liquid surface~\cite{KP1}.

For the general evolutionary problem, we just use the initial loss (${\rm TL}_{I}$) and equation loss (${\rm TL}_{F}$). We choose the first-order rational soliton (also called the lump solution) of KP-I equation~\cite{KP2} as the initial data-set:
\bee\label{solu1}
q(x,y,t)=\frac{F(x,y,t)}{G(x,y,t)},
\ene
where
\begin{equation}\label{FG}
\begin{aligned}
F(x,y,t) =& 256y^2x^2 + 512y^4 + 320y^2 + 512a^4y^4- 12288y^2tx +147456y^2t^2   +32x^4+ 110592x^2t^2 \\
   & + 10616832t^4 - 1769472t^3x +2304xt  - 48x^2 - 27648t^2  - 3072x^3t + 18,\\
G(x,y,t) =& (4x^2 + 16y^2 - 192xt+2304t^2+1)^2.
\end{aligned}
\end{equation}
The rational soliton decays more slowly than the usual solitons with exponential decline~\cite{soliton}.

We here choose $(x,y,t)\in [-2,2]\times[-2,2]\times[-0.05,0.05]$ as the training domain. The 2,000 sample points ($N_I=2,000$) are selected in $(x,y)\in [-2,2]\times[-2,2]$ for the initial data set, and 5,000 sample points ($N_S=5,000$) are selected in the solution region $(x,y,t)\in [-2,2]\times[-2,2]\times[-0.05,0.05]$ by Latin Hypercube Sampling strategy \cite{Latin}. Notice that we here consider the free boundary conditions. We use a 6-layer fully connected neural network with 40 neurons per layer and a hyperbolic tangent activation function to approximate the solution $q(x,y,t)$. 10,000 steps Adam and 20,000 steps L-BFGS optimizations \cite{BFGS} are used in the training processes.
Notice that in each step of the L-BFGS optimization, the program is stopped at
  $ |{\rm loss}(n)-{\rm loss}(n-1)|/{\rm max}(|{\rm loss}(n)|,|{\rm loss}(n-1)|,1)<1.0\times {\rm np.finfo(float).eps}$,

\begin{figure}[!t]
\begin{center}
{\scalebox{0.65}[0.6]{\includegraphics{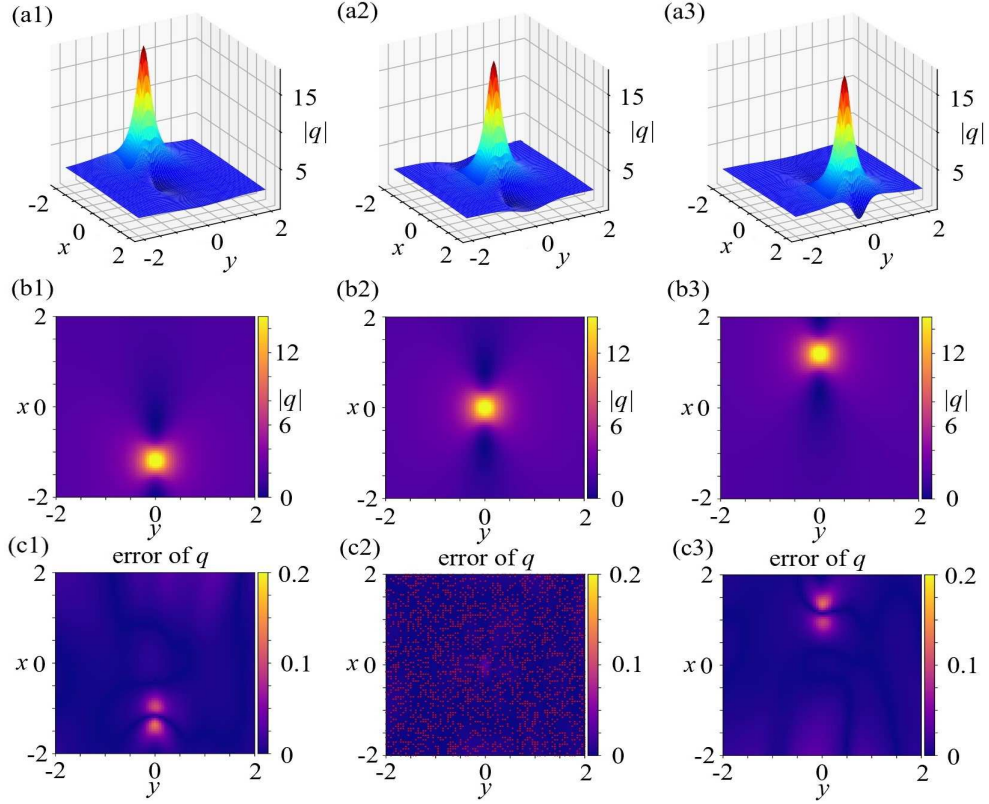}}}
\end{center}
\par
\vspace{-0.1in}
\caption{\small Data-driven rational soliton (\ref{solu1}) of the KP-I equation (\ref{KP-I}). (a1-a3) the approximated solutions at $t=-0.05,\,0,\,0.05$. (b1-b3) density profiles of the approximated solutions at $t=-0.05,\,0,\,0.05$. (c1-c3) The absolute errors of the approximated solutions at $t=-0.05,\,0,\,0.05$. The relative $\mathbb{L}^2-$norm errors of $q(x,t)$, $u(x,t)$ and $v(x,t)$, respectively, are $2.878\cdot 10^{-3}$, $2.878\cdot 10^{-3}$, $0$. The training time is 1910.19s. The red dots in (c2) denote all sampling points of initial data set. }
\label{fig2-rw}
\end{figure}

 Figs.~\ref{fig2-rw}(a1-a3, b1-b3) show that the approximated solution is shifting at the same speed along the $x$-axis. Figs.~\ref{fig2-rw}(c1-c3) display that the PINN method can simulate rational soliton in the high-dimensional condition. The relative $\mathbb{L}^2-$norm errors of $q(x,t)$, $u(x,t)$ and $v(x,t)$ are $2.878\cdot 10^{-3}$, $2.878\cdot 10^{-3}$, $0$, respectively. The learning times is 1910.19s by using a Lenovo notebook with a 2.6GHz six-cores, i7 processor and a RTX2060 graphics processor.

\subsection{Data-driven rational solitons of the (2+1)-D spin-NLS equation}

In this subsection, we will consider the rational solitons of the (2+1)-dimensional spin-nonlinear Schr\"odinger (spin-NLS) equation:
\bee\label{NLSs}
iq_t+\alpha_1q_{xx}+\alpha_2q_{yy}+\alpha_3q_{xy}-\alpha_4|q|^2q=0,
\ene
where $q=q(x,y,t)$ denotes the complex envelope field, and the four real free parameters $\alpha_1, \alpha_2, \alpha_3$ and $\alpha_4$. Let $\xi=x+ky$, Eq.~(\ref{NLSs}) can be transformed into the (1+1)-dimensional NLS equation:
\bee\label{NLS1}
iq_t+(\alpha_1+\alpha_2k^2+\alpha_3k)q_{\xi\xi}-\alpha_4|q|^2q=0.
\ene
In the following, we will consider the first- and second-order rational solitons  of Eq.~(\ref{NLSs}).

\subsubsection{Data-driven first-order W-shaped rational solitons}

The first-order W-shaped rational soliton of Eq.~(\ref{NLSs}) can be obtained by Darboux transform~\cite{NLS1}:
\bee\label{solu2}
q(x,y,t)=\Big(\frac{4(1-2i\alpha_4t)}{4k^2y^2+4\alpha_4^2t^2+8kxy+4x^2+1}-1\Big)e^{-i\alpha_4t},\quad
k=-\frac{\alpha_3\mp\sqrt{\alpha_3^2-4\alpha_2(\alpha_1+\frac{1}{2}\alpha_4)}}{2\alpha_2}.
\ene

We use a 6-layer fully connected neural network with 40 neurons per layer and hyperbolic tangent activation function to approximate the first-order rational soliton (\ref{solu2}) of Eq.~(\ref{NLSs}) . In order to ensure the accuracy of initial data, 1,000 initial sampling points ($N_I=1,000$) are randomly chosen in $(x,y)\in [-10,10]\times[-10,10]$ at $t=0$ and $\alpha_1=0.4$, $\alpha_2=1$, $\alpha_3=-1$, $\alpha_4=-2$ from the rational soliton (\ref{solu2}). And 5,000 training sampling points ($N_S=5,000$) are chosen in $(x,y,t)\in [-10,10]\times[-10,10]\times[-5,5]$ by Latin Hypercube Sampling strategy \cite{Latin}.

 Fig.~\ref{fig3-1rw} exhibits the approximated solutions and absolute errors at $t=0,\,1,\,5$. The relative $\mathbb{L}^2-$norm errors of $q(x,t)$, $u(x,t)$ and $v(x,t)$ are $2.449\cdot 10^{-2}$, $1.940\cdot 10^{-1}$, and $2.238\cdot 10^{-1}$, respectively.
In particular, it follows from Figs~\ref{fig3-1rw}(a1-a3) that when the time increases from $0$ to $5$, the highest amplitude of the  W-shaped rational soliton gradually decreases, and approaches the constant $1$, that is, the W-shaped rational soliton finally degenerates the plane wave.

\begin{figure}[!t]
\begin{center}
{\scalebox{0.65}[0.6]{\includegraphics{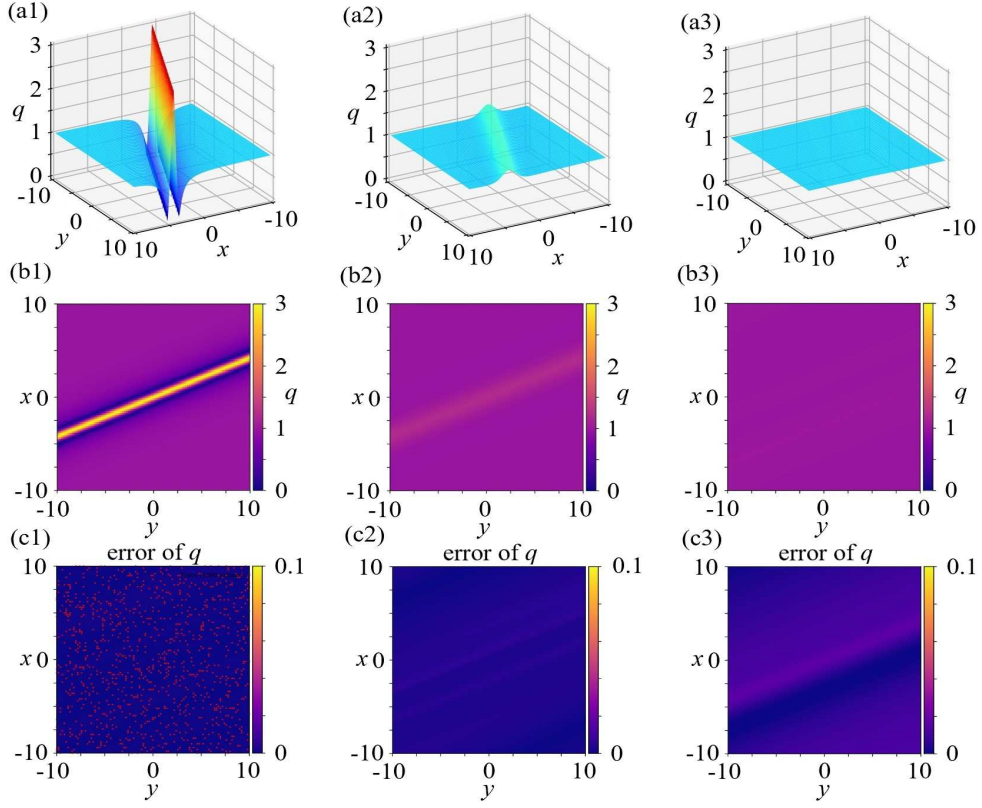}}}
\end{center}
\par
\vspace{-0.1in}
\caption{\protect\small Data-driven 1-order rational soliton (\ref{solu2}) of the spin-NLS equation (\ref{NLSs}). (a1-a3) The neural network solutions at $t=0,\,1,\,5$, (b1-b3) the density profiles of the neural network solutions at $t=0,\,1,\,5$, (c1-c3) the absolute errors of the neural network solutions at $t=0,\,1,\,5$. The red dots in (c1) represent all initial sampling points. The relative $\mathbb{L}^2-$norm errors of $q(x,t)$, $u(x,t)$ and $v(x,t)$, respectively, are  $2.449\cdot 10^{-2}$, $1.940\cdot 10^{-1}$, $2.238\cdot 10^{-1}$. The training time is $2115.12s$. }
\label{fig3-1rw}
\end{figure}

\subsubsection{Data-driven second-order W-shaped rational solitons}

The second-order rational solitons of (\ref{NLSs}) can be obtained~\cite{NLS1}
\bee\label{solu3}
q(x,y,t)=\Big(1+\frac{G_{21}-i\alpha_4G_{22}t}{H_2}\Big)e^{-i\alpha_4t},
\ene
where
\begin{equation}\label{GH}
\begin{aligned}
G_{21} =& - 192k^4y^4 - 1152k^2\alpha_4^2t^2y^2 - 960\alpha_4^4t^4 - 768k^3xy^3 - 2304k\alpha_4^2t^2xy - 1152k^2x^2y^2 \\
        & - 1152\alpha_4^2t^2x^2 - 768kx^3y - 288k^2y^2 - 864\alpha_4^2t^2 - 192x^4 - 576kxy - 288x^2 + 36,\\
G_{22} =& - 384k^4y^4 - 768k^2\alpha_4^2t^2y^2 - 384\alpha_4^4t^4 - 1536k^3xy^3 - 1536k\alpha_4^2t^2xy - 2304k^2x^2y^2 \\
        & - 768\alpha_4^2t^2x^2 - 1536kx^3y + 576k^2y^2 - 192\alpha_4^2t^2 - 384x^4 + 1152kxy + 576x^2 + 360,\\
H_2 =& 64k^6y^6 + 192k^4\alpha_4^2t^2y^4 + 192k^2\alpha_4^4t^4y^2 + 64t^6\alpha_4^6 + 384k^5xy^5 + 768k^3\alpha_4^2t^2xy^3 + 384k\alpha_4^4t^4xy \\
        & + 960k^4x^2y^4 + 1152k^2\alpha_4^2t^2x^2y^2  + 192\alpha_4^4t^4x^2 + 1280k^3x^3y^3 + 768k\alpha_4^2t^2x^3y + 48k^4y^4+ 108x^2 \\
        & - 288k^2\alpha_4^2t^2y^2 + 960k^2x^4y^2 + 432\alpha_4^4t^4 + 192\alpha_4^2t^2x^4  + 192k^3xy^3 - 576k\alpha_4^2t^2xy+ 384kx^5y \\
        & + 288k^2x^2y^2 - 288\alpha_4^2t^2x^2 + 64x^6 + 192kx^3y + 108k^2y^2 + 396\alpha_4^2t^2 + 48x^4+ 108x^2 + 9,
\end{aligned}
\end{equation}
where $k$ still satisfies (\ref{solu2}).

The solution (\ref{solu3}) would be approximated by a 9-layer fully connected neural network with 20 neurons per layer. And the hyperbolic tangent activation function would be used to approximate the nonlinearity of (\ref{solu3}). 1,000 initial sampling points ($N_I=1,000$) are randomly chosen in $(x,y)\in [-5,5]\times[-5,5]$ at $t=0$, and 5,000 training sampling points ($N_S=5,000$) are chosen in $(x,y,t)\in [-5,5]\times[-5,5]\times[-2,2]$ by Latin Hypercube Sampling strategy \cite{Latin}. In this example, we assume $\alpha_1=0.4$, $\alpha_2=1$, $\alpha_3=-1$, $\alpha_4=-2$.

\begin{figure}[!t]
\begin{center}
{\scalebox{0.73}[0.7]{\includegraphics{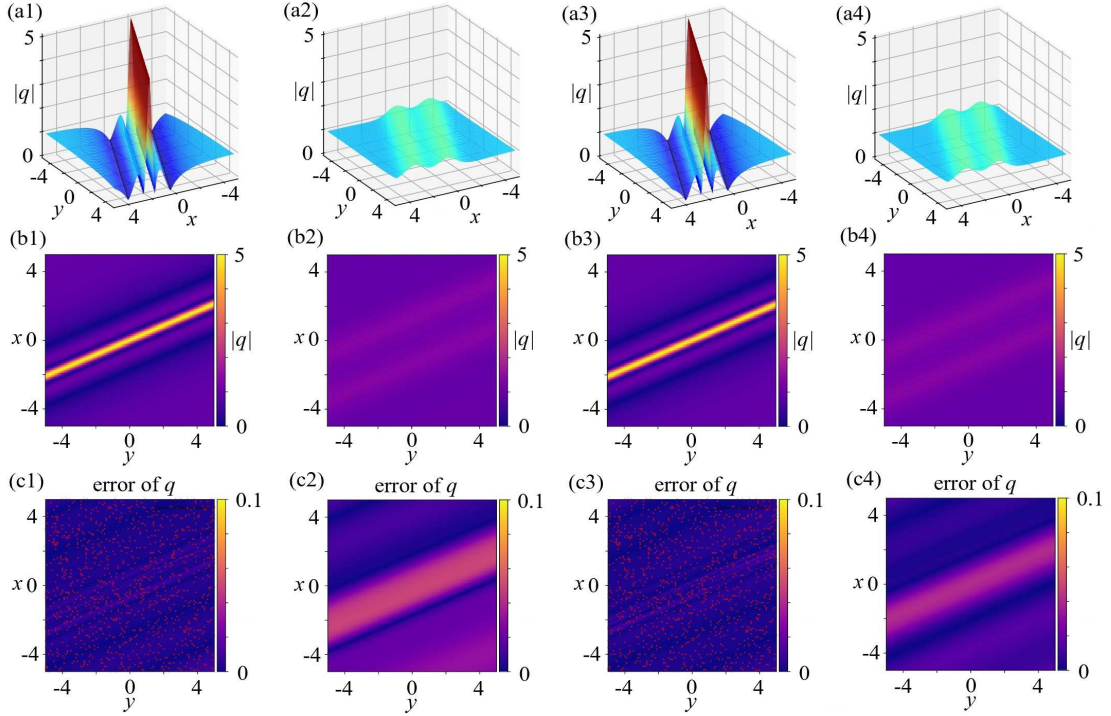}}}
\end{center}
\par
\vspace{-0.1in}
\caption{\protect\small Data-driven 2-order rational soliton (\ref{solu3}) of the (2+1)-D spin-NLS equation (\ref{NLSs}). (a1, a2) and (a3, a4) the approximated solutions arising from the unperturbated and perturbated ($2\%$) training data related to the 2-order rational soliton initial data, respectively; (b1, b2, b3, b4) the corresponding density profiles of (a1, a2, a3, a4); (c1-c4) the absolute errors of the approximated solutions at $t=0,\,1$. The red dots in (c1, c3) represent all initial sampling points. The relative $\mathbb{L}^2-$norm errors of $q(x,t)$, $u(x,t)$ and $v(x,t)$, respectively, are (unperturbated) $3.537\cdot 10^{-2}$, $1.552\cdot 10^{-1}$, $1.164\cdot 10^{-1}$, (perturbated) $3.180\cdot 10^{-2}$, $1.461\cdot 10^{-1}$, $1.086\cdot 10^{-1}$, . The training time of these two cases are $1326.92s$ and $1203.36s$, respectively. }
\label{fig4-2rw}
\end{figure}

The training results are exhibited in Fig.~\ref{fig4-2rw}. We divide the training processes into the unperturbated case (see, Figs.~\ref{fig4-2rw}(a1, a2, b1, b2, c1, c2)) and perturbated case (see, Figs.~\ref{fig4-2rw}(a3, a4, b3, b4, c3, c4)). The initial data set will be added $2\%$ noise in the perturbated case. The relative $\mathbb{L}^2-$norm errors of $q(x,t)$, real part $u(x,t)$ and imaginary part $v(x,t)$, respectively, are (unperturbated) $3.537\cdot 10^{-2}$, $1.552\cdot 10^{-1}$, $1.164\cdot 10^{-1}$, (perturbated) $3.180\cdot 10^{-2}$, $1.461\cdot 10^{-1}$, $1.086\cdot 10^{-1}$, . The training times of these two cases are $1326.92s$ and $1203.36s$, respectively. Moreover, the same wave phenomena are also found similarly to the case of the data-driven first-order rational solitons in Sec. 2.3.1.

\section{The inverse problems of higher-dimensional PDEs}

\begin{figure}[!t]
\begin{center}
{\scalebox{0.75}[0.75]{\hspace{0.25in}\includegraphics{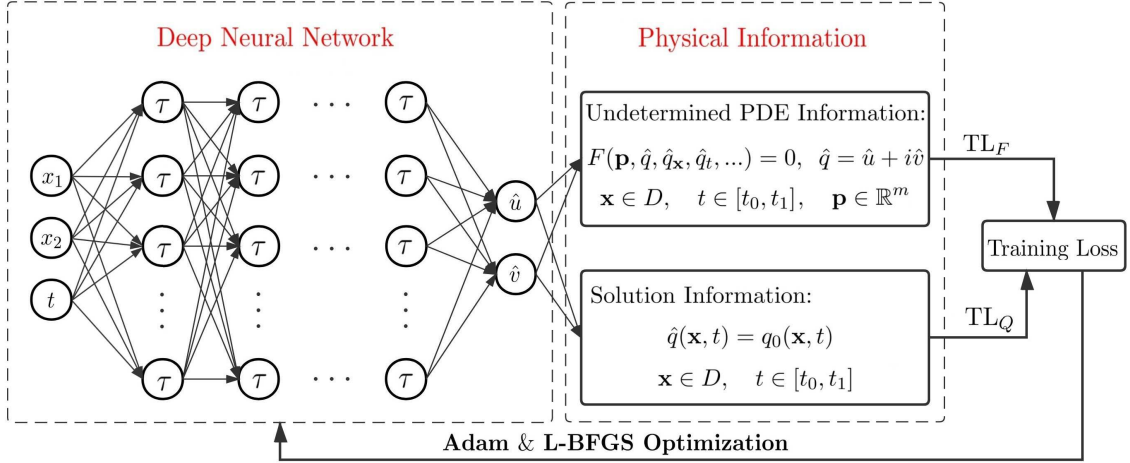}}}
\end{center}
\par
\vspace{-0.05in}
\caption{\small The PINN scheme solving the inverse problem (\ref{prob2}), where ${\cal T}$ denotes the activation function (e.g., ${\cal T}=\tanh(\cdot)$).}
\label{fig5-DNN2}
\end{figure}

In this section, we apply the PINNs deep learning method to solve the inverse problem (\ref{prob2}) (i.e., discovering the parameters of the unknown equation $F(\bp,q,q_{\bx},q_t,...)=0$). Firstly, we would like to briefly introduce the PINN scheme for the inverse problem in the high-dimensional case (see Fig.~\ref{fig5-DNN2}). The loss function contains two parts: the solution loss (TL$_Q$) and undetermined-equation loss (TL$_F$). And a hyper-parameters $\lambda_Q$ is added in the solution loss term (TL$_Q$) to balance the influence of solution loss and undetermined-equation loss during the optimization processes:
\bee
{\rm TL}=\lambda_Q {\rm TL}_Q+{\rm TL}_F,
\ene
where
\bee\begin{array}{l}
 {\rm TL}_{Q}=\displaystyle\frac{1}{N_Q}\sum_{j=1}^{N_Q}\left[\left(\hat u(\bx_Q^j, t_Q^j)-u(\bx_Q^j,t_Q^j)\right)^2
 +\left(\hat v(\bx_Q^j, t_Q^j)-v(\bx_Q^j,t_Q^j)\right)^2\right],\v\\
 {\rm TL}_{F}=\displaystyle\frac{1}{N_F}\sum_{j=1}^{N_F}\left(F_u^2(\bx_F^j, t_F^j)+F_v^2(\bx_F^j, t_F^j)\right).
 \end{array}
\ene
As we did in the previous section, we divide ${\rm TL}_Q$ and ${\rm TL}_F$ into the real and imaginary parts, where $\{\bx_Q^j,\, t_Q^j,\, u(\bx_Q^j,t_Q^j),\, v(\bx_Q^j,t_Q^j)\}_{j=1}^{N_Q}$ denote the solution data-set, and $\{\bx_F^j,\, t_F^j\}_{j=1}^{N_F}$ represent the collocation points of PINN $F(q,q_x,q_t,...)=F_u(q,q_x,q_t,...)+iF_v(q,q_x,q_t,...)$ within a spatio-temporal region $(\bx,t)\in [-L, L]^2\times [t_0, t_1]$. This loss function will be trained by the Adam and L-BFGS optimization algorithm~\cite{BFGS}.

\subsection{Data-driven parameter discovery for the (2+1)-D KP-I equation}

In this part, we would like to consider the parameters discovery for the (2+1)-D KP-I equation through the PINNs deep learning method. We rewrite the original KP-I equation (\ref{KP-I}) as the extended form
\bee\label{KP2}
(q_t+aqq_x+bq_{xxx}+cqq_{xxx})_x-q_{yy}=0,
\ene
where $a,\, b$ and $c$ represent the unknown parameters, and the perturbation term $qq_{xxx}$ is added in the equation to verify the validity of the algorithm.

We will learn this equation in two cases. In the first case, the original KP-I equation will be learned without the perturbation term, i.e., $c=0$. And in the second case, the KP-I equation will be learned with the above-mentioned perturbation term. Because the solution data of KP-I equation are all real, so we initialize the neural network with one output to represent the solution of KP-I equation. We use a 5-hidden layer with 40 neurons per layer to approximate the solution $q(x,y,t)$. The hyper-parameters $\lambda_Q$ is set to 20 to increase the capability of solution approximation. 10,000 steps Adam and 50,000 steps L-BFGS optimization algorithm are used to train the model. The training data are generated by the rational soliton (\ref{solu1}), witch is an exact solution of (2+1)-D KP-I equation. The training domain is selected as $(x,y,t)\in [-2,2]\times[-2,2]\times[-0.05,0.05]$, and 10,000 sampling points are randomly chosen from this area.

\begin{table}[!t]
	\centering
    \setlength{\tabcolsep}{8pt}
    \renewcommand{\arraystretch}{1.4}
	\caption{Data-driven parameter discovery of the KP-I equation (\ref{KP-I}). The training times of four cases are 2050.80s, 1963.42s, 1960.73s and 2305.53s. \vspace{0.05in}}
	\begin{tabular}{ccccccc} \hline\hline
	Case & $a$ & error of $a$ & $b$ & error of $b$ & $c$ & error of $c$  \\  \hline
  Exact & $-6$ & 0 & 1 & 0 & 0 & 0  \\
  Case 1 (no noise) & $-6.03505$ & 3.50$\times10^{-2}$ & 1.00871 & 8.71$\times10^{-3}$ & 0 & 0   \\
  Case 1 ($2\%$ noise) & $-5.94384$ & 5.62$\times10^{-2}$ & 0.97842 & 2.17$\times10^{-2}$ & 0 & 0   \\
  Case 2 (no noise) & $-6.21099$ & 2.11$\times10^{-1}$ & 1.05717 & 5.72$\times10^{-2}$ & -0.00071 & $7.13\times10^{-4}$   \\
  Case 2 ($2\%$ noise) & $-5.88795$ & $1.12\times10^{-1}$ & 0.95398 & $4.60\times10^{-2}$ & 0.00122 & $1.22\times10^{-3}$   \\
   \hline\hline
\end{tabular}
	\label{KP-learn}
\end{table}

Table~\ref{KP-learn} shows the data-driven parameter discovery of the  KP-I equation. The maximum relative error of undetermined parameters are $7.73\times10^{-3}$, $1.04\times10^{-2}$, $5.41\times10^{-2}$ and $4.82\times10^{-2}$ in above four cases. We can find that the PINN scheme can discover the unknown parameters successfully in the different cases. Fig.~\ref{fig6-learnKP} displays the shapes and absolute errors of approximated rational solitons. In Figs.~\ref{fig6-learnKP}(a1-a3), the approximate solution propagates along the $x$-axis. And the relative $\mathbb{L}^2-$norm errors of $q(x,y,t)$, are $1.91\cdot 10^{-2}$, $1.48\cdot 10^{-2}$, $1.87\cdot 10^{-2}$ and $1.66\cdot 10^{-2}$, respectively.

\begin{figure}[!t]
\begin{center}
{\scalebox{0.65}[0.6]{\hspace{0.25in}\includegraphics{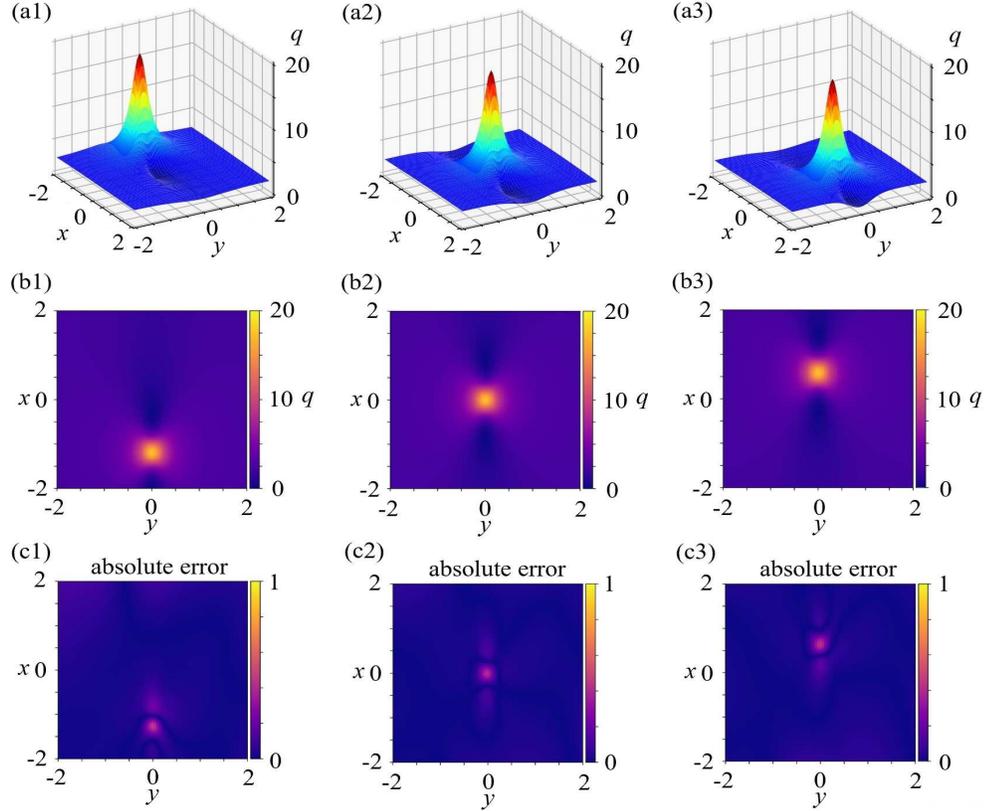}}}
\end{center}
\par
\vspace{-0.15in}
\caption{\small Data-driven inverse problem of the KP-I equation (\ref{KP-I}). (a1-a3) Three profiles of approximate solution at $t=-0.05,\,0,\,0.02$. (b1-b3) The density profiles of approximate solution at $t=-0.05,\,0,\,0.02$. (c1-c3) The absolute errors between approximate and exact solutions. The $\mathbb{L}^2-$norm errors of $q(x,y,t)$, are $1.91\cdot 10^{-2}$, $1.48\cdot 10^{-2}$, $1.87\cdot 10^{-2}$ and $1.66\cdot 10^{-2}$, respectively.}
\label{fig6-learnKP}
\end{figure}

\subsection{Data-driven parameter discovery for the coupled NLS-cmKdV equations}

For the solution $q(x,y,t)$ of KP-I equation (\ref{KP-I}),  the following transform~\cite{KP2}
\bee\label{Trans}
q(x,y,t)=2|r(x,y,t)|^2,
\ene
where $r(x,y,t)$ is a complex function of the variables $x,y,t$, can change the $(2+1)$-D KP-I equation (\ref{KP-I}) into the
(1+1)-D NLS equation:
\bee\label{NLS}
ir_y-r_{xx}-2|r|^2r=0,
\ene
and the (1+1)-D focusing complex mKdV (cmKdV) equation:
\bee\label{mKdV}
r_t+4r_{xxx}+24|r|^2r_x=0.
\ene

In this subsection, we will consider the data-driven parameter discovery problem of the above-mentioned NLS and cmKdV equations via the data set of KP-I equation (i.e. we try to learn the unknown parameters of NLS and cmKdV equations using the solution data set of KP-I equation). In order to achieve this goal, a transform loss (TL$_T$) will be added in the loss function \cite{PDE8}:
\bee\label{transloss}
 {\rm TL}_{T}=\displaystyle\frac{1}{N_T}\sum_{j=1}^{N_T}\left[\hat{q}(\bx_T^j, t_T^j)-2\left(\hat{u}(\bx_T^j, t_T^j)^2+\hat{v}(\bx_T^j, t_T^j)^2\right)\right]^2,
\ene
where $\{\bx_T^j,\, t_T^j\}_{j=1}^{N_T}$ represent the collocation points of transform (\ref{Trans}).  In this example, a neural network with three outputs is used to approximate the $q$, $u$ and $v$ ($u={\rm Re}(r)$, $v={\rm Im}(r)$), resepctively. They will share the same neural network parameters. We consider this problem by testing the data-driven parameter discovery of the coupled 
NLS-cmKdV equations from the following two cases:

\begin{itemize}

\item {} In case A, we assume that the unknown coupled NLS-cmKdV equations are:
    \bee\label{sys1}
    \left\{\begin{array}{l}
    ir_y+ar_{xx}-2|r|^2r+crr_x=0,\quad \vspace{0.1in}\\
    r_t+br_{xxx}+24|r|^2r_x+dr_x^2=0,
    \end{array}\right.
    \ene
    where the vector parameter $\bp=(a,b,c,d)$ contains the four unknown parameters, and two perturbed quadratic terms are added in the original equations to test the robustness of this scheme.

\item {} In case B, the unknown coupled NLS-cmKdV equations are taken as:
    \bee\label{sys2}
    \left\{\begin{array}{l}
    ir_y-r_{xx}+a|r|^2r+crr_x=0,\quad \vspace{0.1in}\\
    r_t+4r_{xxx}+b|r|^2r_x+dr_x^2=0,
    \end{array}\right.
    \ene
   where the vector parameter $\bp=(a,b,c,d)$ contains the four unknown parameters in the cubic nonlinear terms and perturbed quadratic terms.

\end{itemize}

A 6-layer fully connected network with 40 neurons per hidden layer will be used in Cases A and B. Both two cases will be trained by 10,000 steps and 20,000 steps L-BFGS optimization algorithm. 20,000 sampling points are randomly selected in $(x,y,t)\in [-2,2]\times[-2,2]\times[-0.05,0.05]$. We still use the rational soliton (\ref{solu1}) as the training data set.

Table~\ref{NLSmKdV} exhibits the training results of the coupled NLS-cmKdV equations. It is clear that the parameters can be learned well in Cases A and B. The maximum absolute errors in Cases A and B are 7.65$\times10^{-4}$, 2.46$\times10^{-1}$ (without noise) and 3.17$\times10^{-3}$, 4.85$\times10^{-2}$ (with $2\%$ noise), respectively. The perturbed nonlinear terms have the effect on the discovery of nonlinear terms in the original equation, and the absolute values of coefficients of nonlinear terms become larger than high-order derivative terms. The $\mathbb{L}^2-$norm errors of data set $q$ are 4.59$\times10^{-3}$ (case A without noise), 3.29$\times10^{-3}$ (case A with $2\%$ noise), 1.06$\times10^{-2}$ (case B without noise), 1.10$\times10^{-2}$ (case A with $2\%$ noise), respectively. The FNN can approximate this data set really well. But the perturbed terms weaken the accuracies of the approximations.

\begin{table}[!t]
	\centering
    \setlength{\tabcolsep}{15pt}
    \renewcommand{\arraystretch}{1.4}
	\caption{Data-driven parameter discovery of the coupled NLS-cmKdV equations (\ref{sys1},\,\ref{sys2}) via the transform (\ref{Trans}): $a, b, c, d$, and their errors, as well as the training times. \vspace{0.05in}}
	\begin{tabular}{cccccc} \hline\hline
	Case & $a$  & $b$ & $c$  & $d$  & \\  \hline
    Exact of A & -1 & 4 & 0 & 0 &  \\
  A (no noise) & -1.00036 & 3.99994  & -0.00071 & -0.00077 &  \\
  A ($2\%$ noise) & -0.99935 & 3.99683 & -0.00014 & -0.00007 &  \\
    Exact of B & -2 & 24 & 0 & 0 &  \\
   B (no noise) & -1.97689 & 23.75477  & -0.00066 & -0.00093 &  \\
   B ($2\%$ noise) & -1.97442  & 23.73410 & -0.00049 &  0.00010 & \\
    \hline\hline
  Case & error of $a$ & error of $b$ & error of $c$ & error of $d$ & time \\  \hline
    A (no noise) & 3.58$\times10^{-4}$ & 6.00$\times10^{-5}$ & 7.06$\times10^{-4}$ & 7.65$\times10^{-4}$ & 2146.21s \\
  A ($2\%$ noise) & 6.51$\times10^{-4}$ & 3.17$\times10^{-3}$ & 1.38$\times10^{-4}$ & 6.80$\times10^{-5}$ & 2479.20s \\
   B (no noise) & 2.31$\times10^{-2}$ & 2.46$\times10^{-1}$ & 6.33$\times10^{-4}$ & 9.27$\times10^{-4}$ & 2610.67s \\
   B ($2\%$ noise) & 4.85$\times10^{-2}$ & 1.08$\times10^{-2}$ & 3.27$\times10^{-4}$ & 3.80$\times10^{-4}$ & 2624.10s \\
    \hline\hline
	\end{tabular}
	\label{NLSmKdV}
\end{table}

\subsection{Data-driven parameter discovery in the (2+1)-D spin-NLS equation}

Here we would like to study the data-driven parameter discovery of the (2+1)-D spin-NLS equation (\ref{NLSs}). 
We will use the data-set generated by the 1-order and 2-order rational solitons given by Eqs.~(\ref{solu2})-(\ref{solu3}) to discover the original equation.  We add a derivative-nonlinear perturbed term to verify the robustness of this scheme. 
The training equation (i.e., (2+1)-D generalized spin-NLS equation) can be written as:
\bee\label{NLS3}
iq_t+\alpha_1q_{xx}+\alpha_2q_{yy}+\alpha_3q_{xy}-\alpha_4|q|^2q+cqq_x=0,
\ene
where $\alpha_1=0.4$, $\alpha_2=1$, $\alpha_3=-1$ are fixed, and the real-valued parameters $\alpha_4$ and $c$ will be learned and excluded during training processes. We set the initial values of $\alpha_4,\, c$ to be $0$ (Notice that the correct values of $\alpha_4,\, c$ should be $-2$ and $0$, respectively). We use the data-set given by Eqs.~(\ref{solu2})-(\ref{solu3}) with $\alpha_4=-2$ in Cases A and B, respectively. A 6-layer neural network with 40 neurons per layer is used to approximate the solution data. 10,000 sampling points are randomly selected in $(x,y,t)\in [-5,5]\times[-5,5]\times[-2,2]$, and the residual points of TL$_F$ are set to the same 10,000 sampling points.
 
\begin{table}[!t]
	\centering
    \setlength{\tabcolsep}{15pt}
    \renewcommand{\arraystretch}{1.4}
	\caption{Data-driven parameter discovery of the (2+1)-D spin-NLS equation (\ref{NLS3}): $\alpha_4,\, c$, and their errors, as well as training times. \vspace{0.05in}}
	\begin{tabular}{cccccc} \hline\hline
	Case & $\alpha_4$ & error of $\alpha_4$  & $c$ & error of $c$ & time \\  \hline
    Exact of A & -2 & 0 & 0 & 0 &  \\
   A (no noise) & -1.99998 & 1.90$\times10^{-5}$ & -0.00047 & 4.65$\times10^{-4}$ & 1906.72s \\
   A ($2\%$ noise) & -1.99964 & 3.59$\times10^{-4}$ & 0.00060 & 5.95$\times10^{-4}$ & 1839.14s \\
    Exact of B & -2 & 0 & 0 & 0 &  \\
   B (no noise) & -1.99990 & 1.02$\times10^{-4}$  & 0.00121 & 1.21$\times10^{-3}$ & 2393.06s \\
   B ($2\%$ noise) & -2.00037  & 3.72$\times10^{-4}$ & 0.00038 &  3.84$\times10^{-4}$ & 2910.89s \\
    \hline\hline
	\end{tabular}
	\label{SNLS}
\end{table}

We can find that the parameters $\alpha_4, \, c$ of the previous nonlinear term and perturbed term can be discovered 
with the different data-set without noise or with a $2\%$ noise (see Table~\ref{SNLS}). The relative $\mathbb{L}^2$ norm of $q(x,y,t)$, real part $u(x,y,t)$ and imaginary part $v(x,y,t)$ are 5.04$\times 10^{-2}$, 4.16$\times 10^{-2}$, 4.40$\times 10^{-2}$
for the case A without noise, 4.52$\times 10^{-2}$, 4.61$\times 10^{-2}$, 6.15$\times 10^{-2}$ for the case A with $2\%$ noise,  3.20$\times 10^{-1}$, 3.60$\times 10^{-1}$, 2.30$\times 10^{-1}$ for the case B without noise, 7.85$\times 10^{-1}$, 7.88$\times 10^{-1}$, 3.58$\times 10^{-1}$ for the case B with $2\%$ noise. These result show that low frequency data-set are easier than high-frequency.

\section{Conclusions and discussions}

In conclusion, we have explored the data-driven rational solitons and parameter discovery of the (2+1)-D KP equation and
(2+1)-D spin-NLS equation via the deep nueral networks learning method. Especially, in the inverse problems, we use the PINNs to directly study the data-driven discovery parameters of the KP-I equation, and to investigate the parameters of NLS and cmKdV equations through the transform (\ref{Trans}). Moreover, the data-driven parameter discovery of the (2+1)-D spin-NLS equation is investigated. The method can also be extended to other high-dimensional nonlinear wave equations.




\v \noindent {\bf Acknowledgments} \v

This work is supported by  the National Natural Science Foundation of
China (Nos. 11925108 and 11731014).





\end{document}